\documentclass[preprint,12pt]{elsarticle}

\usepackage{amssymb}
\usepackage{amsmath}

\journal{Chaos, Solitons and Fractals}

\begin{document}

\begin{frontmatter}



\title{Emergence of continuously varying critical exponents in coupled map lattice as an effect of quenched disorder}


\author[label1]{Priyanka D. Bhoyar\corref{cor1}}
\affiliation[label1]{organization={Department of Physics, Seth Kesarimal  Porwal College},
            city={Kamptee},
            postcode={441001}, 
            state={Maharashtra},
            country={India}}
\cortext[cor1]{pribhoyar@gmail.com}
\author[label2]{Govindan Rangarajan}
\affiliation[label2]{organization={Department of Mathematics, Indian Institute of Science },
            city={Bangalore},
            postcode={560012}, 
            state={Karnataka},
            country={India}}

\author[label3]{Prashant M. Gade\corref{cor2} }
\affiliation[label3]{organization={Department of Physics,RTM Nagpur University },
            city={Nagpur},
            postcode={440033}, 
            state={Maharashtra},
            country={India}}
\cortext[cor2]{prashant.m.gade@gmail.com}
\begin{abstract}
The transition to an absorbing
phase in a spatiotemporal system is a well-investigated nonequilibrium dynamic
transition. The absorbing phase transitions fall into a few universality classes, defined by the critical exponents observed at the critical point. 
We present a coupled map lattice (CML) model
with quenched disorder in the couplings.  In this model, spatial disorders are introduced in the form of asymmetric coupling with a larger coupling ($p$) to a neighbor on the right and a smaller coupling ($1-p$) to the neighbor on the left, for $0 \le p \le0.5$. 
For $p=0$, the system belongs to the directed percolation universality class. For $p>0$, we observe continuously changing critical exponents at the critical point. The order parameter is the fraction of turbulent sites $m(t)$. 
We observe a power-law decay, $m(t) \sim t^{-\delta}$,
at the critical point $\epsilon_c$, where $\epsilon$ is the diffusive coupling parameter. These exponents change continuously and do not match any known universality class in any limit. This could be related to changes in the eigenvalue spectrum of the connectivity matrix as the disorder is introduced. 
\end{abstract}

\begin{graphicalabstract}
\includegraphics{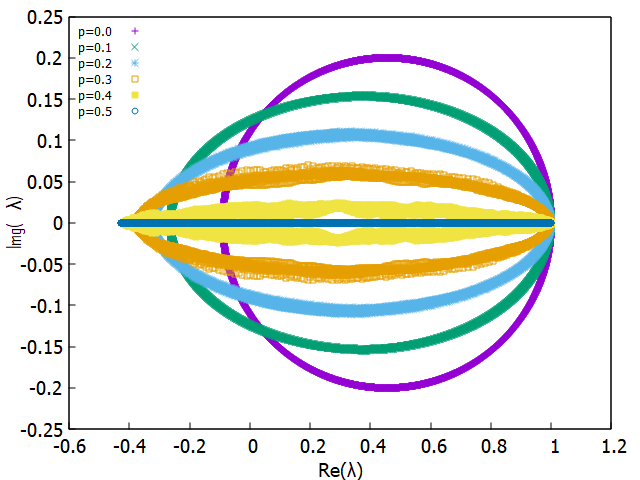}
We consider a connectivity in which all connections are asymmetric. With probability $p$ or ($1-p$), the asymmetry is towards the right or left. The typical eigenvalue spectrum for such connectivity is shown. 
\end{graphicalabstract}

\begin{highlights}
\item {We observe a violation of the universality class in the presence of quenched disorder in the case of a well-known model of the directed percolation universality class.}
\item {The spatial disorder in the form of asymmetric connections in CML leads to non-universal continuously changing critical exponents with a change in the fraction of quenched disorder.} 
\item {A range of power-law decay of order parameter in the absorbing phase is observed with a change in coupling parameter $\epsilon$.  }
\end{highlights}

\end{frontmatter}



\section{Introduction}
\label{sec1}
For many years, researchers have investigated the continuous phase transition. They have been classified into different universality classes. The universality class depends on the dimension and symmetries. Universality is well understood in equilibrium statistical mechanics.  On the other hand, in non-equilibrium systems, it is not as well understood. 
The universality classes are characterized by
a set of critical exponents and scaling functions.
The critical exponents are robust and do not change for a variety of systems in the same universality class. Absorbing phase transitions are the most investigated non-equilibrium transitions.  For absorbing phase transitions,  directed percolation (DP), directed Ising, Manna class etc. are well-investigated classes\cite{henkel2008scaling}. Recently, a new universality class for period-n synchronization has been proposed\cite{joshi2024cellular}. It is fair to say that there are only a handful of known universality classes for absorbing phase transitions in one dimension. 

One of the possible relevant perturbations that could change the universality class is quenched disorder. In some cases, it does not alter the critical behavior at all. In some cases, it leads to a smeared transition. In few cases, it leads to a new universality class. For the contact process, which is usually in the DP class, the quenched disorder may result in the activated scaling universality class. In this case, a logarithmic decay is observed at the critical point and a continuously changing power-law is seen below the critical point\cite{vojta}. Recently, it has been observed that the critical exponents could change with the introduction of quenched disorder for transitions in a model in the multiplicative noise universality class\cite{sabe2024synchronization}. 
In this case, the exponents are superuniversal, but depend on details of the underlying system. 
A change in the value of the critical exponent with a change in system parameters is a violation of the universality class. This could be a signature of the new universality class. It is important because only a few universality classes are known. The universality classes
are differentiated by the precise values of the critical exponents. However, what if the exponent changes continuously {\em{at the critical point}} when a parameter is changed?  
This is different from the Griffiths phase, where
continuously changing critical exponents are observed in a range of values {\em {below}} the
critical point. In the case studied in this work, power-law decay is observed at the critical
point. The exponent changes when we change the system, {\it{i.e.}} on changing the strength of the disorder.

The continuously changing critical exponent has been observed in a few theoretical and experimental cases in the case of equilibrium statistical physics.
The first example is the exactly solvable eight-vertex model(EVM) by Baxter \cite{baxter1972partition}.
The Ashkin-Teller model falls into the same universality class as that of EVM \cite{kadanoff1977connections}. 
The continuous variation in the critical exponent is observed experimentally with the chemical substitution $x$ in $Ru_{2-x}Re_{x}Si_{2}$ single crystals where $0.2 \le x \le 0.6$ \cite{butch2009evolution} and the measurements of light scattering in the non-ionic micellar solution \cite{corti1985critical}. Inayat-Hussain studied the generalization of Lifshitz singularity in the n-vector model with
anisotropy of interaction $\lambda$. For a given set of parameters $d$ (dimension),
$n$ (order parameter), and $\sigma$(interaction range), the critical exponents vary continuously with $\lambda$. They not only differ from the isotropic Lifshitz point usually studied, but also continuously vary with $\lambda$ \cite{inayat1990continuously}.
The critical line between localized Ising magnetic phases in the antiferromagnetic
XYZ spin chain model with quenched bond randomness
is described by infinite randomness fixed points (IRFPs) with continuously varying critical exponents in the
disorder-averaged correlation functions \cite{roberts}.

Suzuki proposed the weak universality (WU) scenario to explain the continuous variation of
exponent \cite{suzuki1974new}. Renormalization group theory allows for continuous variation of
the critical exponents along a marginal direction, keeping the scaling relations invariant. The
critical exponents of the ferromagnetic transition, $\beta$, $\gamma$ and $\nu$
change continuously, however their ratios $\frac{\beta}{\nu}$, $\frac{\gamma}{\nu}$, $\frac{2-\alpha}{\nu}$, and the
exponents $\delta$, and $\eta$ remain invariant \cite{baxter1972partition}.

However, if all exponents vary continuously, the system violates the weak universality hypothesis as well. The critical exponent $\eta$ changes
with the form of the interaction distribution in the case of Ising spin glasses \cite{bernardi1995violation}. Other examples of violation of the weak universality class are
quenched quantum electrodynamics (QED) \cite{kondo1991critical}, ferromagnetic phase 
transition in $(Sm_{1-y}Nd_{y})_{0.52}$ $Sr_{0.48}MnO_3$ single
crystals with $0.5\le y \le1$ \cite{khan2017continuously}.
Within the framework of equilibrium statistical mechanics, Mukherjee $\it{et} \it{al.}$ proposed
a super universality hypothesis (SUH),
suggesting that the scaling functions along the critical line must be identical to those of the base universality class even when all critical exponents vary continuously \cite{mukherjee2023hidden}.

In nonequilibrium systems, a long-range order resulting from local chaos could result in a critical exponent that changes continuously. In the weak coupling domain, the persistence probability decreases algebraically with diffusion coupling strength. It is discovered that the related persistence exponent varies continuously with parameters \cite{lemaitre1999phase}. Another example of continuously changing exponents emerging due to long-range order is that of Levy-flight
type infection spreading into an epidemic process \cite{janssen1999levy}.
In case of a 1D quantum contact process, the critical exponent $\alpha$ changes continuously from its quantum value $0.32$ to its classical value $0.16$
as the strength of the classical fluctuation is varied \cite{jo2021absorbing}.  
Dhar observed self-organized criticality (SOC) can exhibit non-universal scaling depending on system parameters in a two-dimensional model \cite{dhar1993self}.
Noh and Park argued that the long-term memory effect in the generalized paired contact process with diffusion (GPCPD)
acts as a marginal operator that leads to a continuously varying exponent along the phase transition line \cite{noh2004universality}.
This is, to the best of our knowledge, a unique case that exhibits weak universality where the original model belongs to the directed percolation (DP) class.  We note that, although extensively debated, it is suggested that PCPD belongs to the DP universality class \cite{smallenburg2008universality,matte2016persistence}.
Recently, Yi $\it{et}\; \it{al.}$ analyzed the critical properties of
a generalized antiferromagnetic cluster XY model in a transverse magnetic field with algebraically decaying long-range interactions. This article provides a clear example of nonuniversal critical behavior in equilibrium systems \cite{yi2025continuously}. In another work, Delfino investigates the nonuniversal critical behavior of N-coupled two-dimensional Ising models in the presence of quenched disorder. In this case, a line of fixed points in the renormalization group (RG) is observed for any fixed value of N other than 1. The critical exponents vary continuously along this line\cite{delfino2025nonuniversality}.
Saha and Mohanty studied the $q$-state Potts model with non-reciprocal interactions, which drive the system out of equilibrium. They found that for $q = 3$ and $q = 4$, the critical exponents vary continuously, indicating a breakdown of universality due to the non-equilibrium dynamics introduced by asymmetric interactions\cite{saha2024nonreciprocal}.

In this work, we observe a continuous change in critical exponents with a change in the strength of quenched disorder in a nonequilibrium model. We study a model of coupled circle maps that originally belonged to the DP universality class \cite{janaki2003evidence}. The DP universality class is frequently observed, yet is robust under relaxation of several requisites \cite{bhoyar2022robustness}. 
We introduce a spatially quenched disorder in the form of asymmetric connections with probability
$p$ or $1-p$ in either direction.
The quenched disorder is known to alter the critical behavior of a system \cite{vojta}.
Vojta compiled several examples in which a variety of models originally
belonging to the DP universality class changed to the activated scaling class. This class is characterized by a generic power-law decay of the order parameter over a range of parameter values. At the critical point, ultra-slow logarithmic decay is observed.

However, in this work, we observe a power-law decay of the order parameter at the critical point for all values of $p$.
The order parameter saturates for values of $\epsilon$ above the critical point and asymptotically reaches a zero value below the critical point. The decay exponent $\delta$ at the critical point varies continuously with $p$.
Similarly, this case does not belong to the weak universality class as all critical exponents vary continuously. We study the non-universal critical behavior shown by the CML in the following section.

 



\section{Model and simulation}
\label{sec2}
We study the effect of quenched spatial disorder on
the 1D coupled map lattice.
We study the continuous phase transition from the laminar state
to the turbulent state in a well-known DP model studied by Janaki and Sinha \cite{janaki2003evidence}. This model is defined as follows. We associate a variable
$x_i(t)$ with the site $i$ at time $t$.
The time evolution of each site
depends on the state of its nearest neighbor as follows:
$x_{i}(t+1) = (1-\epsilon)f(x_{i}(t))+\frac{\epsilon }{2}[f(x_{i-1}(t))+f(x_{i+1}(t))]\vert_{{\rm{mod}}\; 1}$
where $i=1,2 \dot{...}N$ and $\epsilon$ is the coupling parameter.
The underlying map is a circle map defined as:
$f(x)=x+\omega-\frac{k}{2}{\pi} \sin(2\pi x)$,
where $k$ is the nonlinearity parameter.

The control parameters are fixed at $k=3.2$ and $\omega=0.29$.
The initial conditions $x_{i}(t)$ are randomly chosen in a unit interval.
We label sites as turbulent or laminar depending on their value.
Any site $x_{i}(t)>0.5$ is labeled as turbulent and is laminar otherwise. 
On the lines of \cite{janaki2003evidence}, the fraction
of turbulent sites $m(t)$ is the relevant order parameter in this work. They investigated the transition from chaotic to laminar state and
showed that the model is clearly in the DP class.

The spatial disorder is introduced into the lattice in the form of asymmetric connections.
Let $\epsilon_1=\frac{\epsilon}{2}+0.1$, and  $\epsilon_2=\frac{\epsilon}{2}-0.1$. Let the sites be of type A or B. The evolution rules are given as :\\
Type A sites evolve as $x_{i}(t+1) = (1-\epsilon) f(x_{i}(t))+\epsilon_1 f(x_{i-1}(t))+\epsilon_2 f(x_{i+1}(t))\vert_{{\rm{mod} \;1}}$\\
Type B sites evolve as $x_{i}(t+1) = (1-\epsilon) f(x_{i}(t))+\epsilon_2 f(x_{i-1}(t))+\epsilon_1 f(x_{i+1}(t))\vert_{{\rm{mod} \;1}} $.\\ 
We randomly assign sites to be type A with probability $p$ and type B with probability $1-p$. We vary $p$ between 0-0.5. Due to the left-right symmetry of the lattice, the statistical behavior for $1-p$ is expected to be the same as for $p$ and we do not
study $p>0.5$. 
We simulate a 1D lattice of size $N=2 \times 10^5$. 
The order parameter $m(t)$ is the fraction of turbulent sites at time t. $m(t)$ is computed by averaging over at least $500$ disorder configurations and initial conditions.

\begin{figure}[!ht]%
	\centering
	\begin{minipage}[c]{1\linewidth}
	\includegraphics[width=7.3cm]{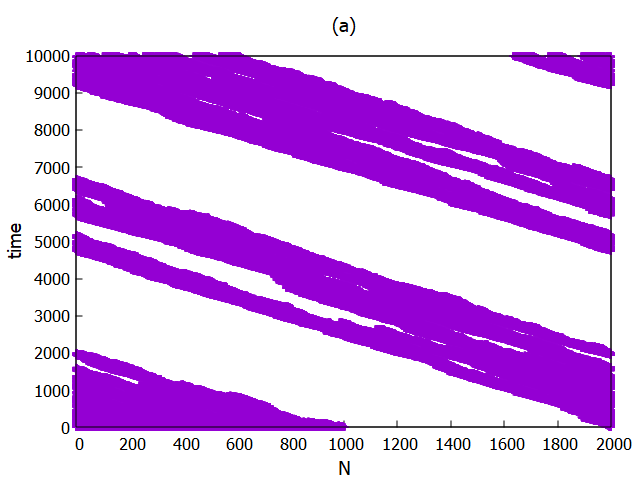}
	\includegraphics[width=7.3cm]{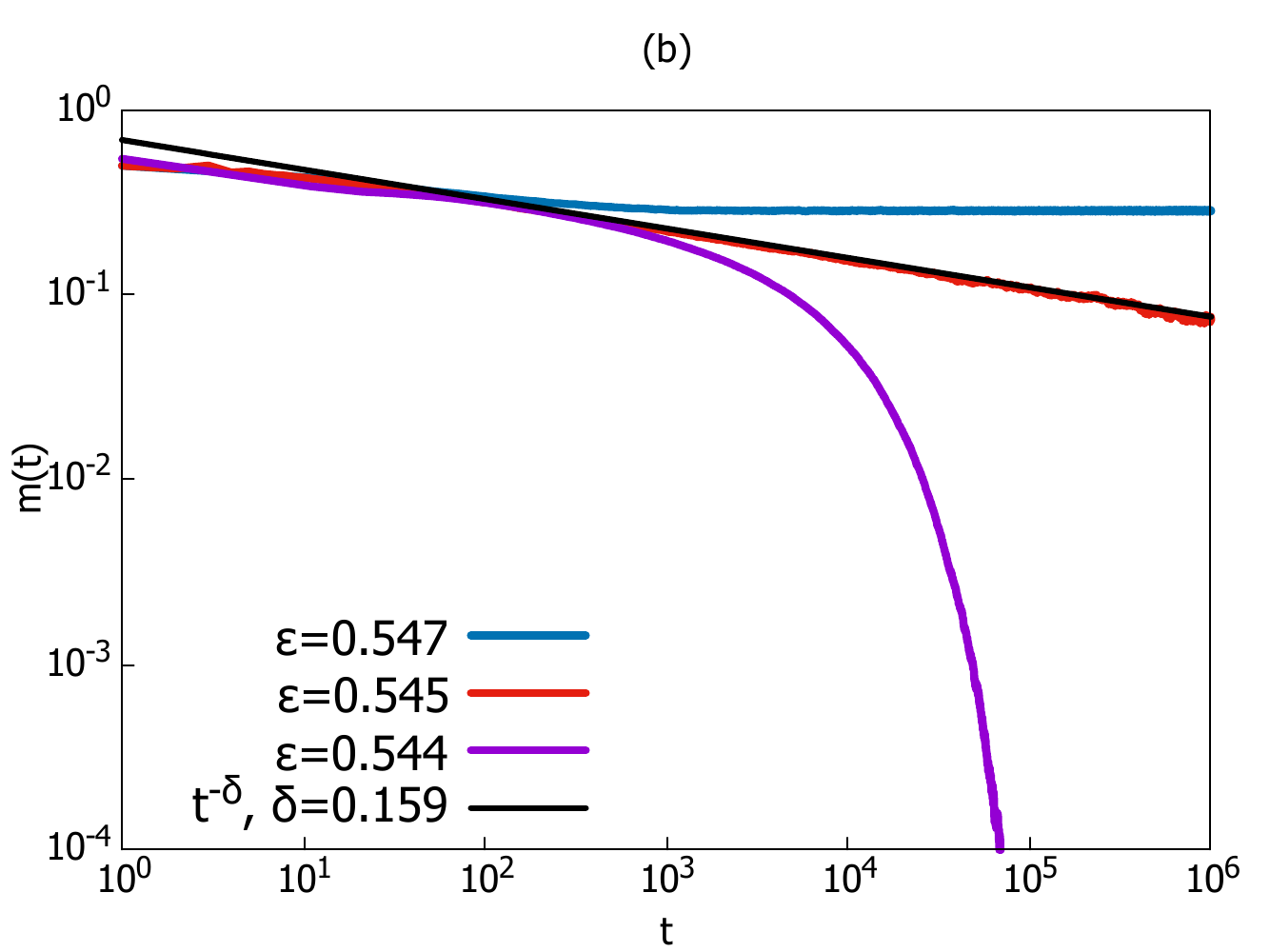}
	\end{minipage}
	\hfill
	\caption{(a) shows the spatiotemporal behavior of the lattice of size $N=2000$ for $p=0$. We randomly activate 5 sites at the centre of the lattice. 
	(b) Shows the evolution of order parameter $m(t)$ with time at $\epsilon > \epsilon_c$, $\epsilon_c$ and 
	$\epsilon < \epsilon_c$ on log-log scale. The system size is $N=2 \times 10^5$ The power-law decay, $m(t)$ $\sim$ $t^{-\delta}$ is observed at $\epsilon_c=0.545$ with $\delta=0.159$.}
	\label{Fig:1}
\end{figure}

The system has completely asymmetric connections for $p=0$.
The DP class is known to be robust under one-sided coupling \cite{tretyakov1997phase}.
The space-time diagram shows the activity completely
driven towards one side (left) (see Fig.\ref{Fig:1}(a)).
The space-time diagram is plotted with five randomly activated sites in the center of the lattice, while
the rest of the sites are at the fixed point,
$x*=\frac{1}{2\pi} \sin^{-1}(2\pi\frac{\omega}{k})$.
The system undergoes a continuous phase transition from
turbulent to laminar state at the critical point $\epsilon_c$=0.545.
The order parameter decays as $m(t) \sim t^{-\delta}$ at $\epsilon_c$=0.545,
where $\delta=0.159$ (see Fig.\ref{Fig:1}(b)).
For $\epsilon > \epsilon_c$, the fraction of turbulent sites $m(\infty)$ reaches a steady state value.
For $\epsilon < \epsilon_c$, the system exponentially approaches an absorbing phase.
The model is expected to be in the DP universality class in this case.

\begin{figure}[!ht]%
	\centering
	\begin{minipage}[c]{1\linewidth}
	\includegraphics[width=7cm]{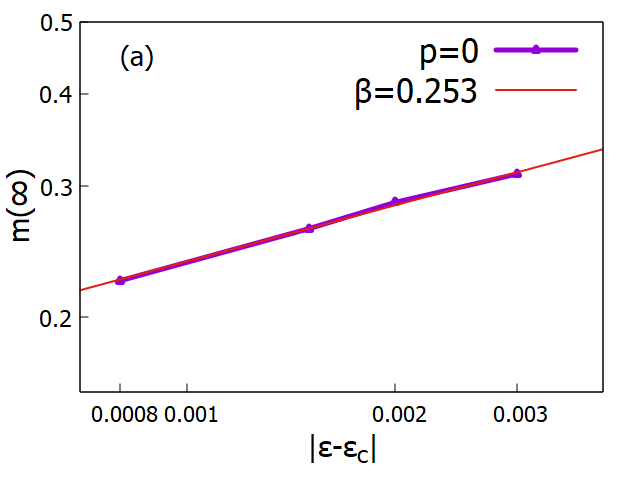}
	\includegraphics[width=7cm]{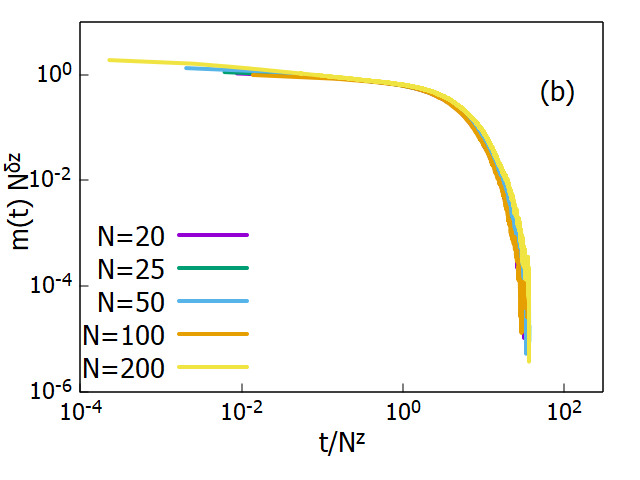}
	\includegraphics[width=7cm]{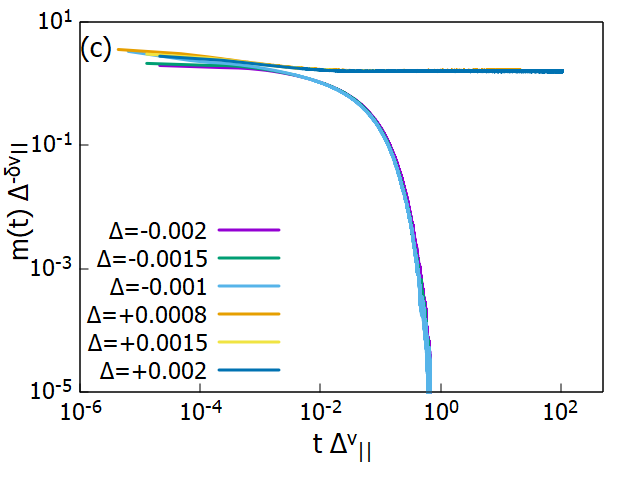}
	\end{minipage}
	\hfill
	\caption{(a) shows the plot of $|\epsilon$-$\epsilon_c|$
	$\it{vs}$  $m(\infty)$ on log-log scale for $p=0$. We obtain $m(\infty) \sim |\epsilon-\epsilon_c|^\beta$  where $\beta$=0.253. (b) Shows the finite-size scaling, $\frac{t}{N^z}$ $\it{vs}$ $m(t) N^{{\delta}z}$ for various sizes of lattice $N=20,25,50,100,200$. The best collapse is obtained for $z$=1.58. and $\delta$=0.159.
    (c) Shows the off-critical scaling, $m(t)\Delta^{-\delta \nu_\parallel}$ $\it{vs}$ $t \Delta^{\nu_{\parallel}}$ for various values of $\Delta$=$\epsilon-\epsilon_c$ in the range [-0.002, 0.002]. The best collapse is obtained for $\nu_\parallel$=1.732 and $\delta=0.159$. All the critical exponents match well with DP values.}
	\label{Fig:2a}
\end{figure}

The DP universality class is characterized by a set of critical exponents: $\delta$, $\beta$, $\nu_\parallel$ and $z$. 
Fig.\ref{Fig:2a} (a) shows the steady-state density  of the turbulent sites $m(\infty)$ as a function of $|\epsilon-\epsilon_c|$. We find
$m(\infty) \sim |\epsilon-\epsilon_c|^ \beta$.
The value obtained for $\beta$ is $0.253$. We perform finite-size scaling and off-critical scaling to determine the critical exponents $z$ and $\nu_{||}$, respectively. We plot $m(t) N^{{\delta}z}$ vs $t/N^{z}$ on the log-log scale in Fig.\ref{Fig:2a}(b). In Fig.\ref{Fig:2a}(c), we plot 
$m(t) \Delta^{-\delta \nu_\parallel}$ vs $t \Delta ^{\nu_\parallel}$. Here $\Delta=|\epsilon-\epsilon_c|$
The best scaling collapse is obtained for $z=1.58$ and $\nu_{||}=1.732$ (see Fig.\ref{Fig:2a}(b) and (c), respectively). The values of these critical exponents match well with the DP values. The activity is perceived in one direction only, with no effect of asymmetry on the exponents.
 
\begin{figure}[!ht]%
	\centering
	\begin{minipage}[c]{1\linewidth}
		\includegraphics[width=7cm]{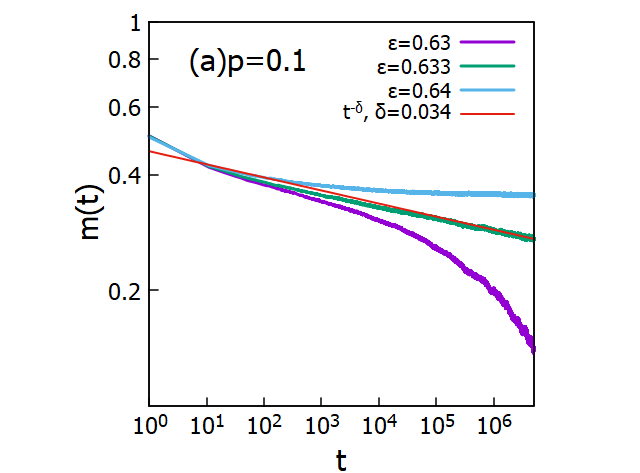}
		\includegraphics[width=7cm]{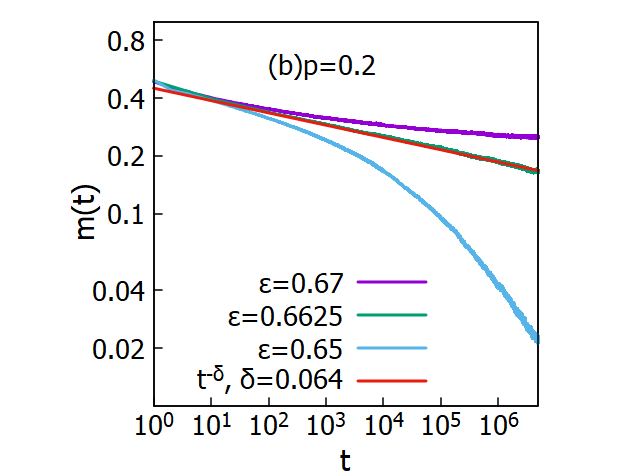}
		\includegraphics[width=7cm]{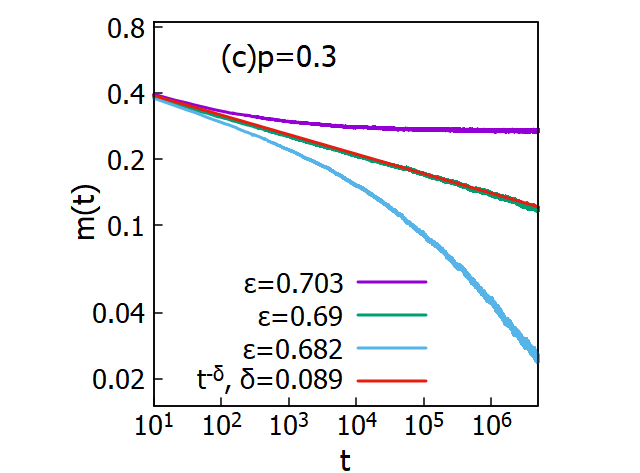}
		\includegraphics[width=7cm]{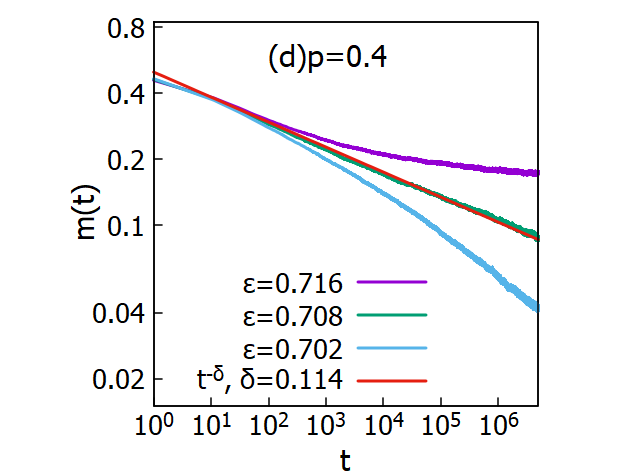}
		\includegraphics[width=7cm]{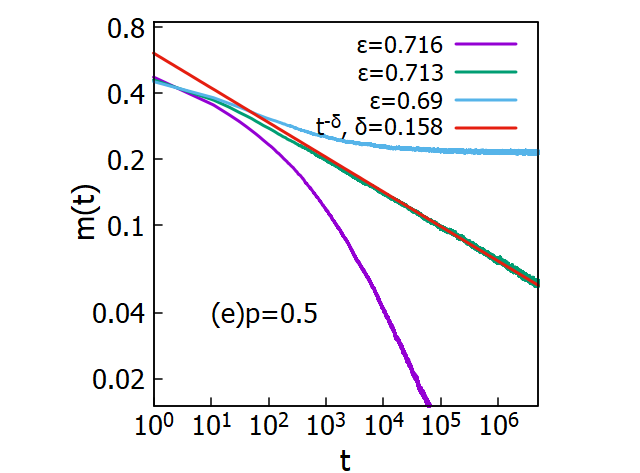}
	\end{minipage}
	\hfill
	\caption{We plot the decay of the order parameter $m(t)$ with time for $\epsilon>\epsilon_c$, $\epsilon$ and $\epsilon<\epsilon_c$(top to bottom). The system size is $N=2\times10^5$. We averaged over 500 disorder configurations and initial conditions. $m(t)$ undergoes power-law decay at $\epsilon_c$, $m(t)\sim t^{-\delta}$. (a) for $p=0.1$, $\epsilon_c$=0.633, $\delta=0.034$
	(b) for $p=0.2$,$\epsilon_c$=0.6625, $\delta=0.064$, (c) for $p=0.3$, $\epsilon_c$=0.69,
	$\delta=0.089$, (d) for $p=0.4$, $\epsilon_c$=0.709, $\delta=0.114$, and (e)for $p=0.5$, $\epsilon_c$=0.713, $\delta=0.158$. A continuous change in decay exponent $\delta$ is observed with change in $p$.}   
	\label{Fig:3}
\end{figure}

For all values of $p>0$, the order parameter undergoes a power-law decay $m(t) \sim t^{-\delta}$ at $\epsilon=\epsilon_c$, similar to DP. As expected, the value of $\epsilon_c$ changes. More importantly,  the decay exponent $\delta$ changes with a change in the value of $p$. Fig.\ref{Fig:3} shows the evolution of the order parameter $m(t)$ for $\epsilon>\epsilon_c$,
$\epsilon=\epsilon_c$, and $\epsilon<\epsilon_c$ for various values of $p$. We obtain the following values. For $p=0.1$, $\epsilon_c$=0.633, $\delta=0.034$, for $p=0.1$, $\epsilon_c$=0.6625, $\delta=0.064$ for $p=0.2$, $\epsilon_c$=0.69, $\delta=0.089$, for $p=0.3$, $\epsilon_c$=0.709, $\delta=0.114$, and for $p=0.5$, $\epsilon_c$=0.713, $\delta=0.158$. These decay exponents do not match any of the known universality classes, except for $\delta$ in case $p=0.5$. In this case, the value of $\delta$ matches with DP. (However, as we see later, the other exponents deviate from the DP class even for $p=0.5$). This universality class is characterized by a continuously changing critical exponent $\delta$ parametrized by the fraction of connections $p$ and $1-p$ in either direction. If the critical exponent varies continuously, it suggests either a new universality class, marginal operators, or non-universal behavior.

\begin{figure}[!ht]%
	\centering
	\begin{minipage}[c]{1\linewidth}
		\includegraphics[width=7cm]{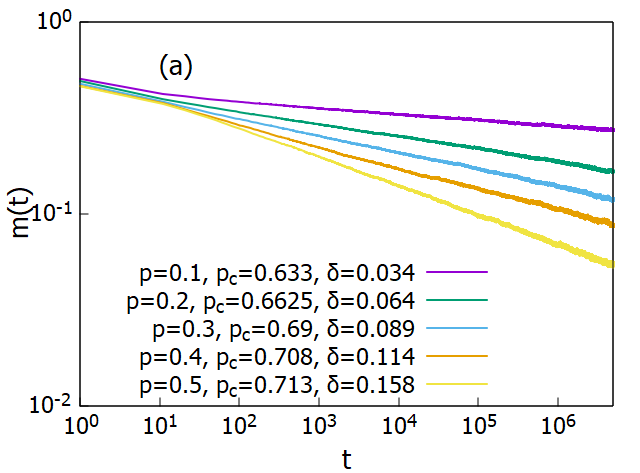}
		\includegraphics[width=7cm]{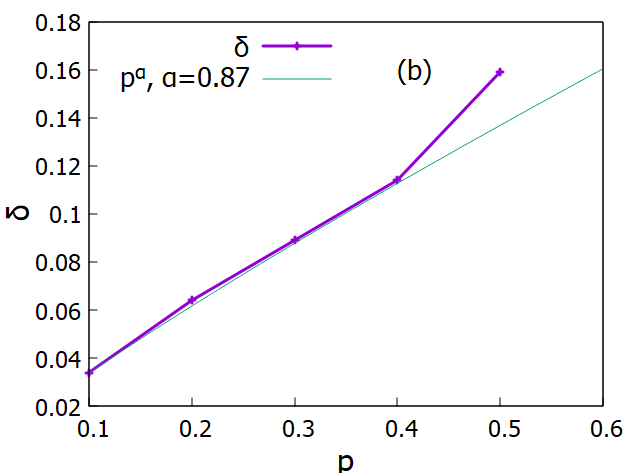}
	\end{minipage}
	\hfill
	\caption{(a) Shows the decay of order parameter for various values of $p=0.1$(top) to $0.5$ (bottom)
	at the critical point $\epsilon_c$. The system size is $N=2\times10^5$. We observe that the model exhibits continuously varying decay exponent $\delta$ with change in $p$. In this case $k=3.2$ and $\omega =0.29$. (b) shows the change in $\delta$ with change in fraction $p$. We find that $\delta \sim p^{\alpha}$ where $\alpha=0.87$.}
	\label{Fig:1a}
\end{figure}

Fig.\ref{Fig:1a}(a) shows the power-law decay for several values of $p$ at the respective critical point. The plot of $\delta$ with $p$ shows that the decay exponent varies almost linearly with $p$. Fig.\ref{Fig:1a}(b)shows that $\delta \sim p^{\alpha}$ where $\alpha=0.87$. Competition between activity in both directions implies that activity does not grow in one direction (unlike $p=0$). 

\begin{figure}[!ht]%
	\centering
	\begin{minipage}[c]{1\linewidth}
		\includegraphics[width=7cm]{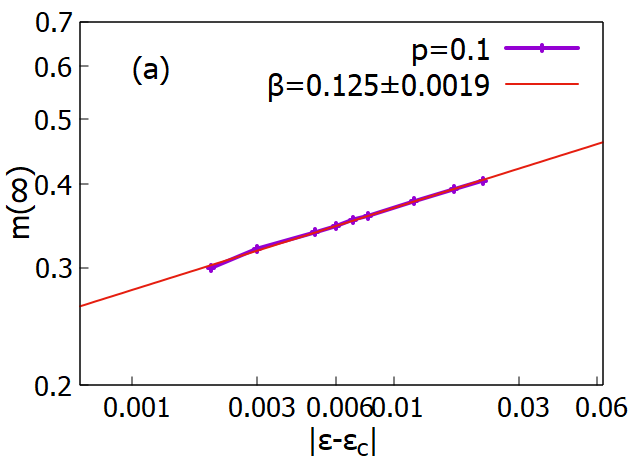}
		\includegraphics[width=7cm]{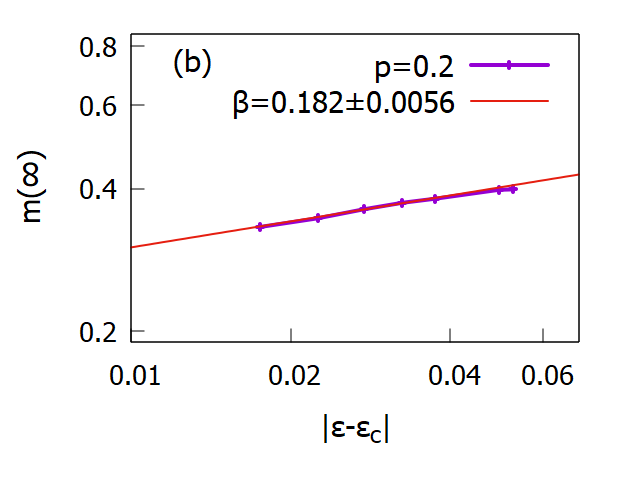}
		\includegraphics[width=7cm]{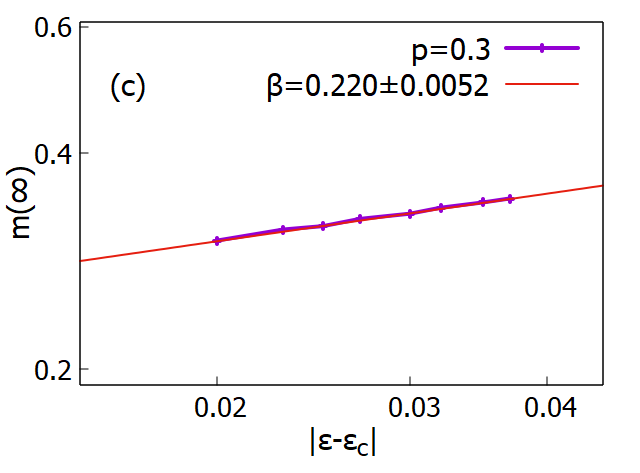}
		\includegraphics[width=7cm]{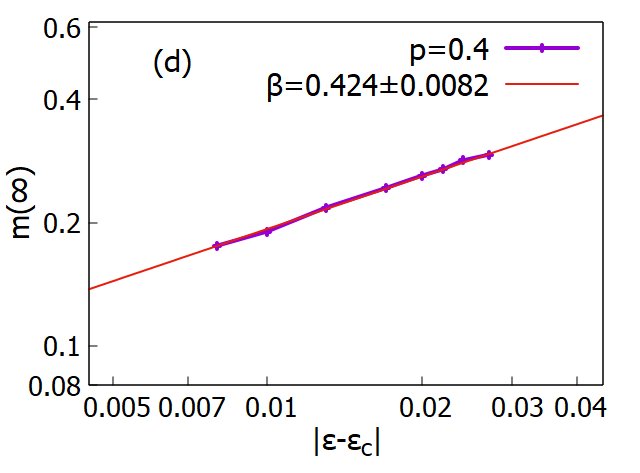}
		\includegraphics[width=7cm]{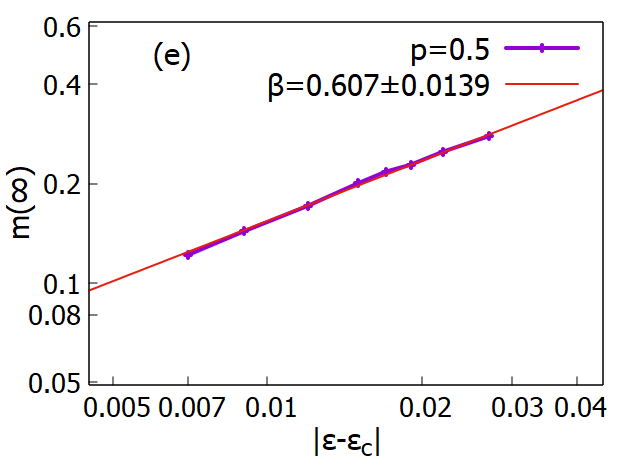}
	\end{minipage}
	\hfill
	\caption{Shows the scaling of asymptotic value of order parameter as $t\rightarrow{\infty}$, $m(\infty)$ with $|\epsilon-\epsilon_c|$. The system size is $2\times10^5$. (a) for $p=0.1$, $\epsilon_c$=0.633, $\beta=0.125\pm0.0019$
	(b) for $p=0.2$,$\epsilon_c$=0.6625, $\beta=0.182\pm0.0056$ (c) for $p=0.3$, $\epsilon_c$=0.69,
	$\beta=0.220\pm0.0052$ (d) for $p=0.4$, $\epsilon_c$=0.709, $\beta=0.424\pm0.0082$ and
	(e) for $p=0.5$, $\epsilon_c$=0.713, $\beta=0.607\pm0.0139$}
	\label{Fig:44}
\end{figure}

In the following section, we obtain the other critical exponents for $p>0$. 
We obtain $\beta$ for $p>0$.
We plot the steady-state density of turbulent sites $m(\infty)$ with respect to $(\epsilon-\epsilon_c)$ for $\epsilon>\epsilon_c$. $m(\infty)\sim (\epsilon-\epsilon_c)^\beta$ for all $p>0$, where $\beta=0.125, 0.182, 0.220, 0.424, 0.607$ for $p=0.1, 0.2, 0.3, 0.4, 0.5$ respectively (See Fig\ref{Fig:44}(a)-(e)). We have estimated errors in the values obtained of $\beta$ using the 'fit' function of
Gnuplot. It uses an implementation of the nonlinear least-squares (NLLS) Marquardt-Levenberg algorithm.
 We observe a continuous increase in the value of $\beta$ with an increase in $p$. The lower value of $\beta$ for the lower value of $p$ indicates that the order parameter increases slowly as the system approaches criticality. This is consistent with the slower decay of the order parameter for smaller values of $p$.

\begin{figure}[!ht]%
	\center
	\begin{minipage}[c]{1\linewidth}
        \includegraphics[width=9cm]{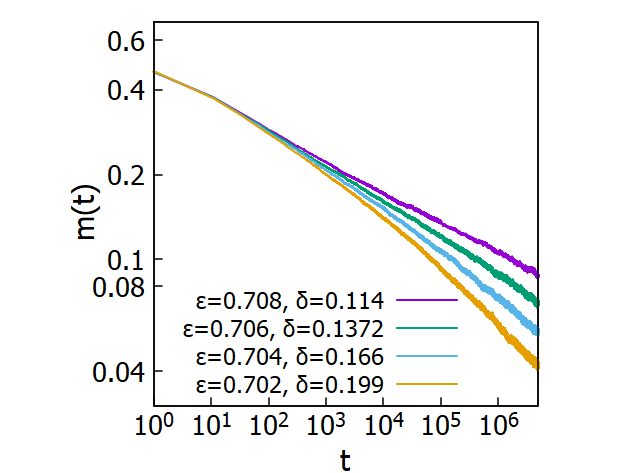}
	\end{minipage}
	\hfill
	\caption{ Shows the algebraic decay of order parameter for several values of $\epsilon$ in the absorbing phase 
	for $p=0.4$. We observe that the order parameter undergoes a range of power-law decay with continuously varying decay exponents in the absorption region. The system size is $2\times10^5$.}
	\label{Fig:11}
\end{figure}

The power-law decay with continuously varying exponent depends on the values of $p$ and $\epsilon$. Fig.\ref{Fig:11} shows several power-laws over a range of $\epsilon$ values in the absorbing phase for $p=0.4$. The decay exponent $\delta$ decreases with an increase in $\epsilon$. The emergence of extended laminar regions having long lifetimes may lead to such behavior near the criticality. The generic power law range drops for lower values of $p$.

\begin{figure}[!ht]%
	\centering
	\begin{minipage}[c]{1\linewidth}
		\includegraphics[width=7cm]{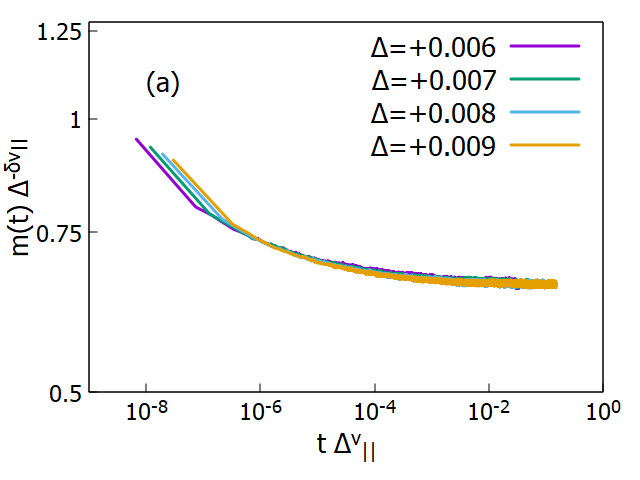}
		\includegraphics[width=7cm]{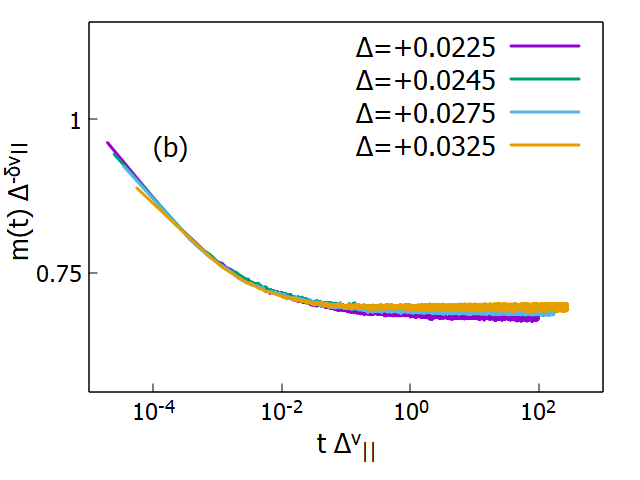}
		\includegraphics[width=7cm]{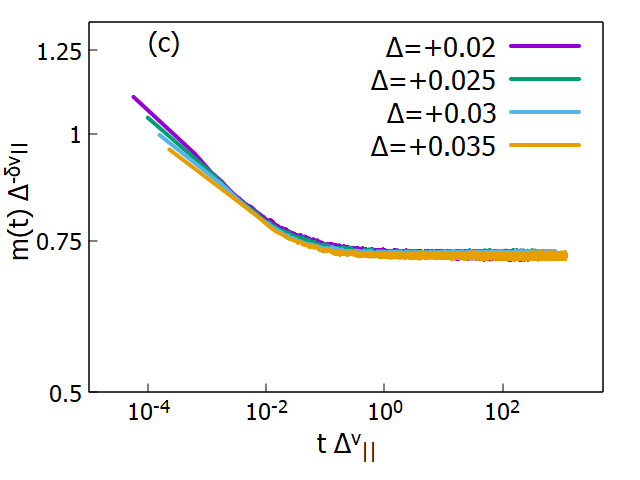}
		\includegraphics[width=7cm]{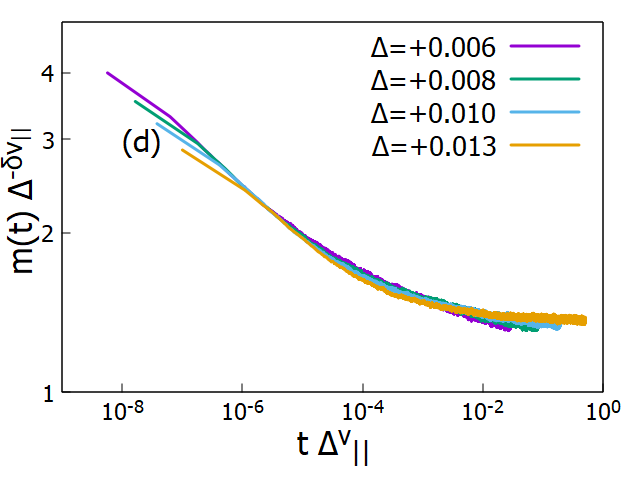}
		\includegraphics[width=7cm]{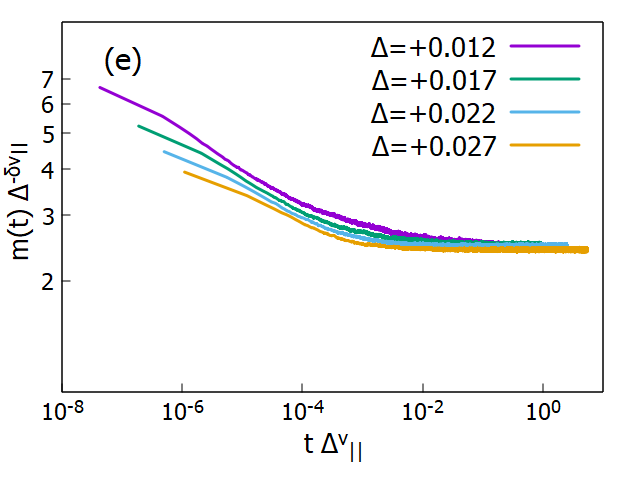}
	\end{minipage}
	\hfill
	\caption{ Shows the off-critical scaling, $t \Delta^{\nu_\parallel}$ vs $m(t) \Delta^{\delta \nu_\parallel}$ on log-log scale for several values of $\epsilon>\epsilon_c$. Here, $\Delta=|\epsilon-\epsilon_c|$. The system size is $2\times10^5$.
    (a) for $p=0.1$, $\epsilon_c=0.633$, $\nu_\parallel=3.67$ (b) for $p=0.2$, $\epsilon_c=0.6625$, $\nu_\parallel=2.86$, (c) for $p=0.3$, $\epsilon_c=0.69$, $\nu_\parallel=2.47$, (d) for $p=0.4$, $\epsilon_c=0.708$, $\nu_\parallel=3.71$, and (e) for $p=0.5$, $\epsilon_c=0.713$, $\nu_\parallel=3.84$.}
	\label{Fig:55}
\end{figure}

We observe spatially heterogeneous scaling behavior in this system for all $p>0$. We plot 
$t|\Delta|^{\nu_\parallel}$ vs $m(t) \Delta^{-\delta \nu_\parallel}$ on log-log scale. Scaling collapse is not observed in the subcritical regime because of the generic power-law decay of the order parameter. The off-critical scaling is done for the supercritical regime, $\Delta = \epsilon>\epsilon_c$ for $p>0$. We obtain the best collapse for the following values of $\nu_\parallel$.
For $p=0.1$, $\nu_\parallel=3.67$, for $p=0.2$, $\nu_\parallel=2.86$, for $p=0.3$, $\nu_\parallel=2.47$, for $p=0.4$, $\nu_\parallel=3.71$, and for $p=0.5$,  $\nu_\parallel=3.84$ (see Fig.\ref{Fig:55}(a)-(e)).
We also expect a similar value of $\nu_\parallel$ from the hyperscaling relation $\nu_\parallel=\frac{\beta}{\delta}$. The hyperscaling relation is valid. Such large values of $\nu_\parallel$ indicate a longer relaxation time.    

\begin{figure}[!ht]%
	\centering
	\begin{minipage}[c]{1\linewidth}
	\includegraphics[width=9cm]{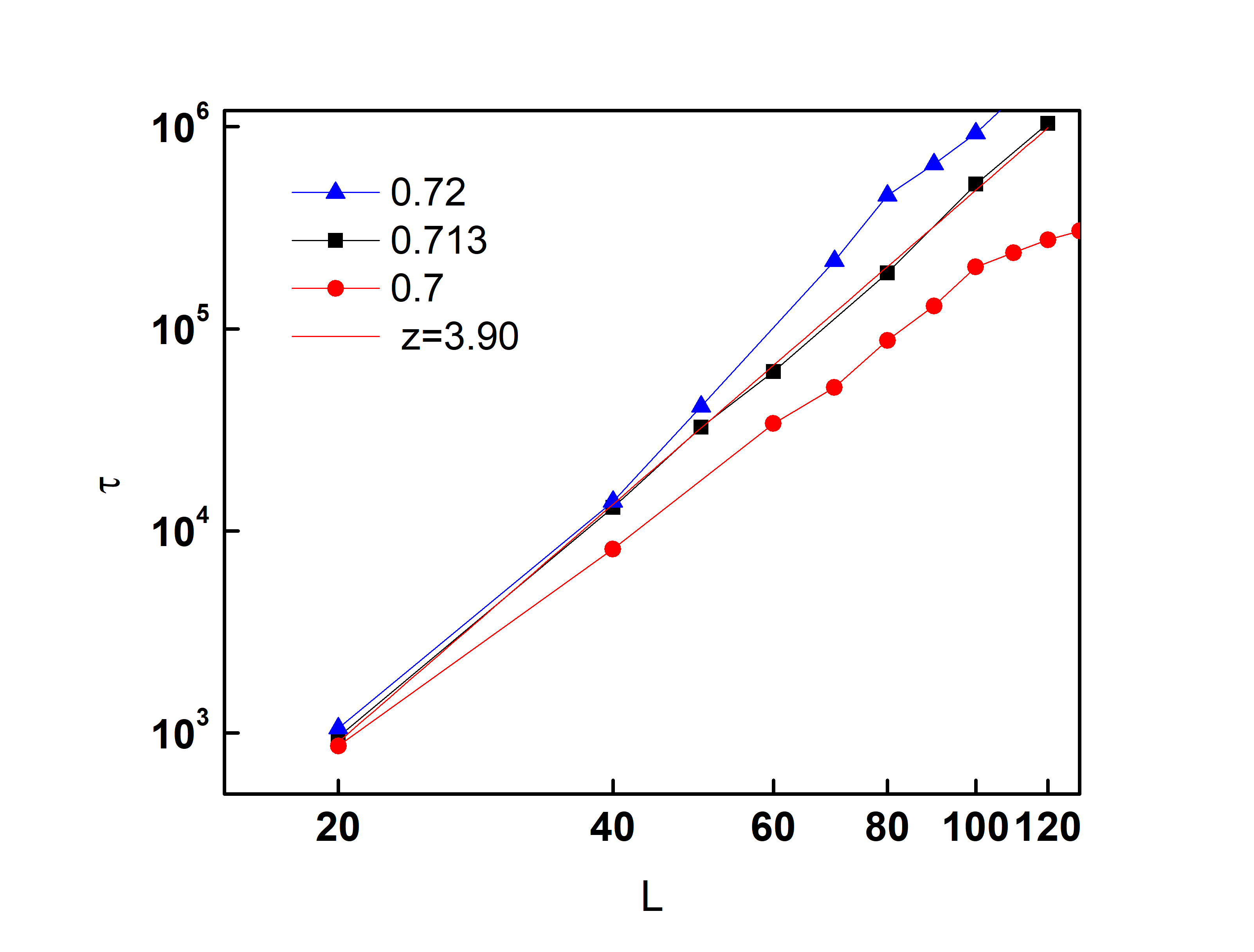}	
	\end{minipage}
	\hfill
	\caption{Shows the plot of the average time $\tau$ taken by system of size $N=20,40,60 \dots$  to reach the absorbing state for $p=0.5$. The power-law is obtained at the critical point with dynamical exponent $z=3.90$.}
	\label{Fig:4}
\end{figure}

\begin{figure}[!ht]%
	\centering
	\begin{minipage}[c]{1\linewidth}                            
	\includegraphics[width=7cm]{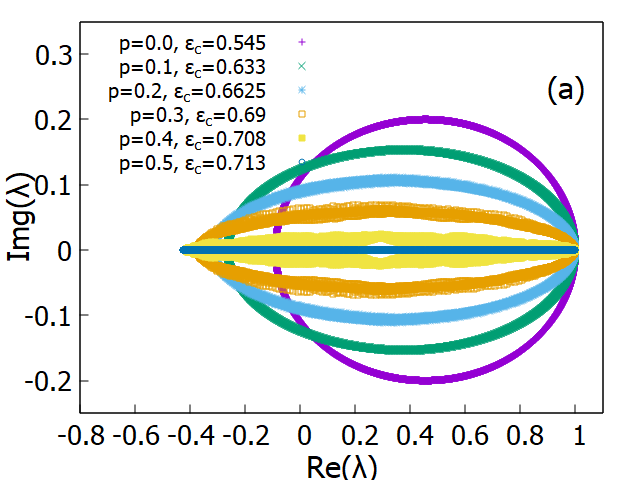}
        \includegraphics[width=7cm]{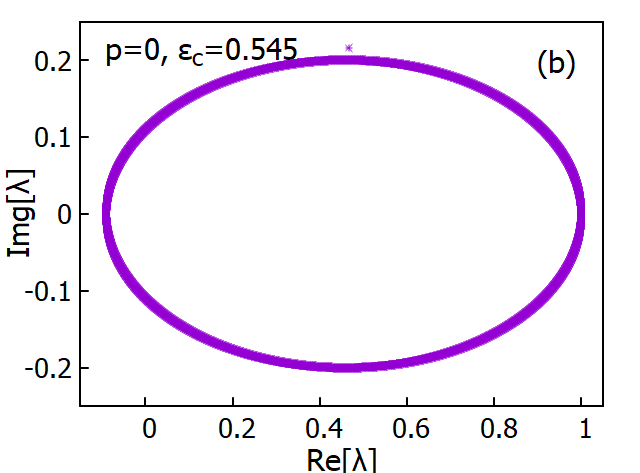}
        \includegraphics[width=7cm]{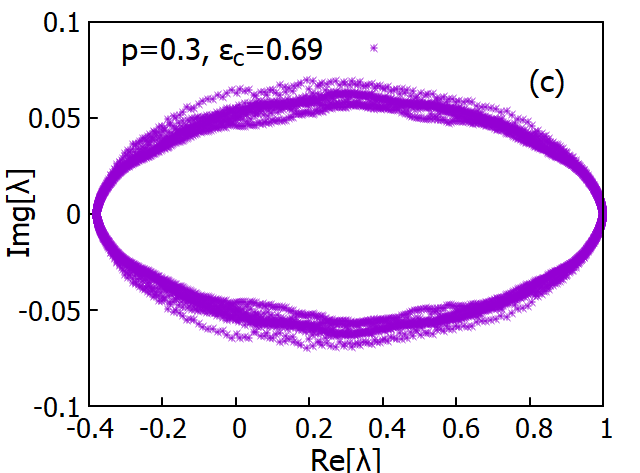}
        \includegraphics[width=7cm]{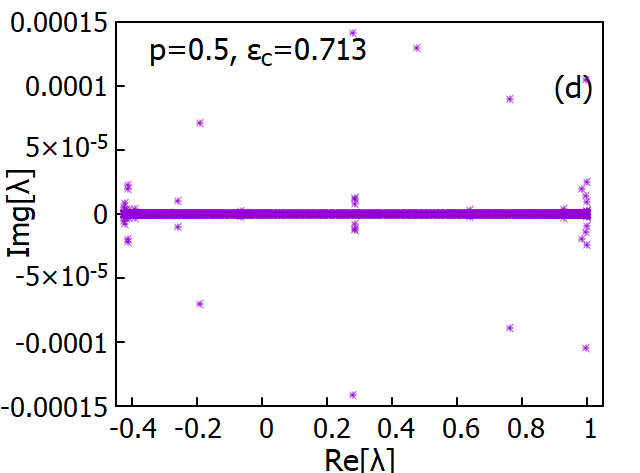}
	\end{minipage}
	\hfill
	\caption{(a) Shows the distribution of eigenvalues in the complex plane for $p=0,0.1,0.2,0.3,0.4,0.5$. The eigen spectra are obtained by exact diagonalization of 2000
	 matrices of size $2000 \times 2000$.	 
	 We observe holes in the spectrum for $p=0-0.4$.
	 The eigenvalues lose their complex nature and become real for $p=0.5$. The magnified view is shown for $p=0$ and $p=0.3
     $ and $p=0.5$ (b) All the eigenvalues are exactly on the perimeter for $p=0$ (c) Some spread around the boundary is observed in the case of $p=0.3$. d) Almost all eigenvalues are real for $p=0.5$.}
\label{Fig:34}
\end{figure}

The dynamical exponent $z$ is expected to be large and hence cannot be obtained for $p>0$ (except for $p=0.5$) because of the ultra-slow decay at the critical point. We could only obtain the dynamical exponent $z$ for $p=0.5$. The average time $\tau$ taken by the system of size $N=20,40,60 \dots$, to reach the absorbing state for different system sizes, starting with a single seed, is shown in Fig.\ref{Fig:4}.
We find a power-law growth at the critical point. The value obtained for the dynamic exponent $z$ is 3.90. This value of $z$ is large. The values of $z$ would be even larger for $0<p<0.5$.
It is beyond our computational limits to find $z$ for other values of $p$. The dynamic exponent $z$ relates the characteristic temporal and the characteristic spatial scale of a system at the 
critical point. Such large values of $z$ show that the characteristic time scale grows faster for a given length scale. (The diffusive system has $z=2$. If $z<2$, the system is considered superdiffusive, as in the case of directed percolation at criticality. In this case, $z>2$ and the system is subdiffusive. This indicates that disturbances move slower than the annihilating random walkers and a longer time is needed for a finite system to reach equilibrium. This may have to do with the fact that the disorder leads to two different competing directions for the defects. Thus the defects move very slowly. This is unlike the case of $p=0$ where they move in the same direction or the case without anisotropy where they spread isotropically.) The slow decay of the order parameter at the critical point is related to slowly moving defects. 
We tabulate all the exponents in Table:\ref{Tab:1}. All the critical exponents $\delta$, $\nu_\parallel$, $\beta$ and $z$ vary continuously with $p$ in our model.
\begin{center}
\begin{table}
\caption{The values of the critical exponents with a change in fraction $p$ }
\label{Tab:1}
\begin{tabular}{| c | c | c | c | c |}
\hline
\textbf{fraction $p$} & \textbf{$\epsilon_c$} & \textbf{$\delta$} & \textbf{$\beta$} & \textbf{$\nu_\parallel$} \\ \hline
0.1 & 0.633 & 0.034 & $0.125$ & 3.67\\ 
0.2 & 0.6625 & 0.064 & $0.182$ & 2.86\\ 
0.3 & 0.69 & 0.089 & $0.220$ & 2.47\\ 
0.4 & 0.708 & 0.114 & $0.424$ & 3.71\\ 
0.5 & 0.713 & 0.158 & $0.607$ & 3.84\\ \hline 
\end{tabular}
\end{table}
\end{center}
The continuous variation of the critical exponents may be related to the changes in the eigenvalue spectrum of the connectivity matrix as the disorder is introduced. We plot the distribution of eigenvalues in a complex plane for each value of $p$ at the critical value of $\epsilon$ in Fig.\ref{Fig:34}. We expect the nature of eigenvalues to be independent of $\epsilon$. The matrices are row stochastic, and the largest eigenvalue is unity. For $p=0$,
the spectrum in the thermodynamic limit given by $(1-\epsilon) + \epsilon_1\exp(i\theta)+\epsilon_2 \exp(-i\theta)= (1-\epsilon) + (\epsilon_1+\epsilon_2) \cos(\theta) + i (\epsilon_1-\epsilon_2) \sin(\theta)$ with $0\le \theta \le 2\pi$. (The parametric equation for an ellipse in the complex plane with zero as centre is given by $a\cos(t)+i \; b\sin(t)$ ($0\le t\le 2\pi$).)Thus, this spectrum is an ellipse in the complex plane with centre $(1-\epsilon)$. All the eigenvalues are exactly on its perimeter.  The eigenvalue spectrum has a hole for $0<p<0.5$. However, the eigenvalues are not on the perimeter of some curves. There is some spread around the boundary. The imaginary component reduces with increasing $p$. For $p=0.5$, the picture changes completely.  The eigenvalues are real for $p=0.5$. Again, they are on a line segment on the real axis, and there is almost no width (imaginary component). Interestingly, although the matrix is still
asymmetric for $p=0.5$, the eigenvalue spectrum 
is real in the thermodynamic limit.   We can say that the connections are "symmetric on average" for $p=0.5$. We note that it is known that a class of non-hermitian matrices can have all real eigenvalues \cite{ashida2020non}. For $p=0.5$,
$\delta$ matches the DP value. However, the values of dynamic exponent
$z$ or exponent $\nu_{\parallel}$ are larger than the expected value
for the DP class.

\section{Summary}
\label{sec3}
We study the effect of spatially quenched disorder on the coupled map lattice model that originally belongs to the directed percolation universality class. The disorder is introduced in the form of asymmetric coupling.
In each variant, a fraction of the connections
$p$ drives the activity in one direction (say right), and the rest of the fraction $1-p$ drives the activity in the other direction (say left). The system is tuned to a critical point similar to that in DP. However,  critical exponents vary continuously and smoothly for non-zero $p$ values. The tussle between these two driving forces leads to a slow power-law for all $p$. The decay exponent $\delta$ changes with $p$, leading to an unknown universality class characterized by continuously changing critical exponents. The exponent $\delta$ again matches the DP class for $p=0.5$.
However, there are notable changes, such as a much higher value of $z$ and $\nu_\parallel$ for $p=0.5$. These values are even higher for smaller values of $p$. Secondly, we also observe a regime of power-law exponents changing over a range of couplings $\epsilon$ in the absorption phase. However, this range decreases with the decrease in the fraction $p$. Thus, we obtain power-law decay with a non-universal exponent that depends on control parameters $p$ and $\epsilon$, unlike any known universality class.



\section{Acknowledgements}
PDB acknowledges Rashtrasant Tukadoji Maharaj Nagpur University for financial assistance (RTMNU/RDC/2024/242).
 \bibliographystyle{elsarticle-num} 
 \bibliography{paper}






\end{document}